\documentclass[%
reprint,
superscriptaddress,
amsmath,amssymb,
prb,
]{revtex4-2}
\usepackage{color}
\usepackage{float}
\usepackage[pdftex]{graphicx}
\usepackage{graphicx}% Include figure files
\usepackage{dcolumn}% Align table columns on decimal point
\usepackage{bm}% bold math
\usepackage{siunitx}
\usepackage{comment}
\usepackage{fixltx2e}
\usepackage{soul}

\usepackage{hyperref}% add hypertext capabilities
\hypersetup{
    bookmarks=true,         % show bookmarks bar?
    unicode=false,          % non-Latin characters in Acrobat’s bookmarks
    pdftoolbar=true,        % show Acrobat’s toolbar?
    pdfmenubar=true,        % show Acrobat’s menu?
    pdffitwindow=false,     % window fit to page when opened
    pdftitle={Charge detection in an array of CMOS quantum dots},    % title
    pdfauthor={Dr. Matias Urdampilleta},     % author
    pdfsubject={},   % subject of the document
    pdfcreator={Dr. Matias Urdampilleta},   % creator of the document
    pdfproducer={},  % producer of the document
    pdfkeywords={,} {} {}, % list of keywords
    pdfnewwindow=true,      % links in new window
    colorlinks=false,       % false: boxed links; true: colored links
    linkcolor=red,          % color of internal links
    citecolor=green,        % color of links to bibliography
    filecolor=magenta,      % color of file links
    urlcolor=cyan           % color of external links
}
\usepackage{filecontents}

\DeclareUnicodeCharacter{2009}{\,} 
\DeclareSIUnit{\belmilliwatt}{Bm}
\DeclareSIUnit{\dBm}{\deci\belmilliwatt}
\newcommand{\ket}[1]{\left| #1 \right>} % for Dirac bras
\newcommand{\bra}[1]{\left< #1 \right|} % for Dirac kets

\begin{document}
\bibliographystyle{apsrev4-1}

\preprint{APS}
\title{Spin-valley coupling anisotropy and noise in CMOS quantum dots}% Force line breaks with \\
\author{Cameron Spence}
\email{cameron.spence@neel.cnrs.fr}
\affiliation{Univ. Grenoble Alpes, CNRS, Grenoble INP, Institut N\'eel, 38402 Grenoble, France}

\author{Bruna Cardoso Paz}
\affiliation{Univ. Grenoble Alpes, CNRS, Grenoble INP, Institut N\'eel, 38402 Grenoble, France}

\author{Bernhard Klemt}
\affiliation{Univ. Grenoble Alpes, CNRS, Grenoble INP, Institut N\'eel, 38402 Grenoble, France}

\author{Emmanuel Chanrion}
\affiliation{Univ. Grenoble Alpes, CNRS, Grenoble INP, Institut N\'eel, 38402 Grenoble, France}

\author{David J. Niegemann}
\affiliation{Univ. Grenoble Alpes, CNRS, Grenoble INP, Institut N\'eel, 38402 Grenoble, France}

\author{Baptiste Jadot}
\affiliation{Univ. Grenoble Alpes, CNRS, Grenoble INP, Institut N\'eel, 38402 Grenoble, France}

\author{Vivien Thiney}
\affiliation{Univ. Grenoble Alpes, CNRS, Grenoble INP, Institut N\'eel, 38402 Grenoble, France}

\author{Benoit Bertrand}
\affiliation{CEA, LETI, Minatec Campus, F-38054 Grenoble, France}
\author{Heimanu Niebojewski}
\affiliation{CEA, LETI, Minatec Campus, F-38054 Grenoble, France}

\author{Pierre-Andr\'e Mortemousque}
\affiliation{CEA, LETI, Minatec Campus, F-38054 Grenoble, France}

\author{Xavier Jehl}
\affiliation{Univ. Grenoble Alpes, CEA, IRIG, 38000 Grenoble, France}

\author{Romain Maurand}
\affiliation{Univ. Grenoble Alpes, CEA, IRIG, 38000 Grenoble, France}

\author{Silvano De Franceschi}
\affiliation{Univ. Grenoble Alpes, CEA, IRIG, 38000 Grenoble, France}

\author{Maud Vinet}
\affiliation{CEA, LETI, Minatec Campus, F-38054 Grenoble, France}

\author{Franck Balestro}
\affiliation{Univ. Grenoble Alpes, CNRS, Grenoble INP, Institut N\'eel, 38402 Grenoble, France}

\author{Christopher B{\"a}uerle}
\affiliation{Univ. Grenoble Alpes, CNRS, Grenoble INP, Institut N\'eel, 38402 Grenoble, France}

\author{Yann-Michel Niquet}
\affiliation{Univ. Grenoble Alpes, CEA, IRIG, 38000 Grenoble, France}

\author{Tristan Meunier}
\affiliation{Univ. Grenoble Alpes, CNRS, Grenoble INP, Institut N\'eel, 38402 Grenoble, France}

\author{Matias Urdampilleta}
\email{matias.urdampilleta@neel.cnrs.fr}
\affiliation{Univ. Grenoble Alpes, CNRS, Grenoble INP, Institut N\'eel, 38402 Grenoble, France}

\date{\today}% It is always \today, today,
\begin{abstract}
One of the main advantages of silicon spin qubits over other solid-state qubits is their inherent scalability and compatibility with the $\SI{300}{mm}$ CMOS fabrication technology that is already widely used in the semiconductor industry, whilst maintaining high readout and gate fidelities.
We demonstrate detection of a single electron spin using energy-selective readout in a CMOS-fabricated nanowire device with an integrated charge detector.
We measure a valley splitting of $\SI{0.3}{meV}$ and $\SI{0.16}{meV}$ in two similar devices.
The anisotropy of the spin-valley mixing is measured and shown to follow the dependence expected from the symmetry of the local confinement, indicating low disorder in the region of the quantum dot.
Finally the charge noise in the spin-valley coupling regime is investigated and found to induce fluctuations in the qubit energy in the range of $\SI{0.6}{GHz}/\sqrt{\SI{}{Hz}}$.
\end{abstract}

\maketitle

%%%%%%%%%%%%%%%%%%%%%%%%%%%%%%%%%%%%%%%%%%%%%%%%%%%%%%%%%%%%%%%%%%%%%%%%
\section{Introduction}

Silicon spin qubits are a highly promising candidate for scalable quantum computation.
High-fidelity spin measurement with integrated charge sensing up to relatively high temperature, fast ESR or EDSR manipulation for one- and two-qubit gates, and long coherence times in isotopically purified silicon all point towards the high potential of the silicon platform for solid-state qubit  \cite{yoneda2018quantum, DzurakReflecto, Huang2019, veldhorst2015two, huang2021}. %cite yoneda2018quantum, DzurakReflecto, Huang2019, veldhorst2015two, huang2021
Additionally, industrial expertise in CMOS semiconductor fabrication provides a clear path towards mass production of nanoscale qubit devices \cite{Veldhorst2017, schaal2019cmos}. %cite Veldhorst2017, schaal2019cmos
Spin readout and manipulation have been demonstrated in CMOS quantum dots with high fidelity \cite{Urdampilleta2019, corna2018, crippa2019}. %Urdampilleta2019, corna2018, crippa2019
However characterization of the spin physics in these types of devices remains an open problem, with indications of local disorder and variability across similar devices \cite{ciriano2021spin}. %Cite ciriano2021spin

In this letter, we present the measurement of spin relaxation in CMOS quantum dots fabricated on a foundry-compatible 300 mm wafer.
We first show how we can rapidly probe the spin using an energy selective readout with more than $90\%$ readout fidelity.
Secondly, we explore the dynamics of spin relaxation in the system and the the coupling of spin and valley states, by measuring the spin lifetime as a function of magnetic field strength and direction.
Finally, we investigate the charge noise in this system by performing spin-valley relaxometry.

%\section{Device design and operation}

\begin{figure}%
\includegraphics[width=\columnwidth]{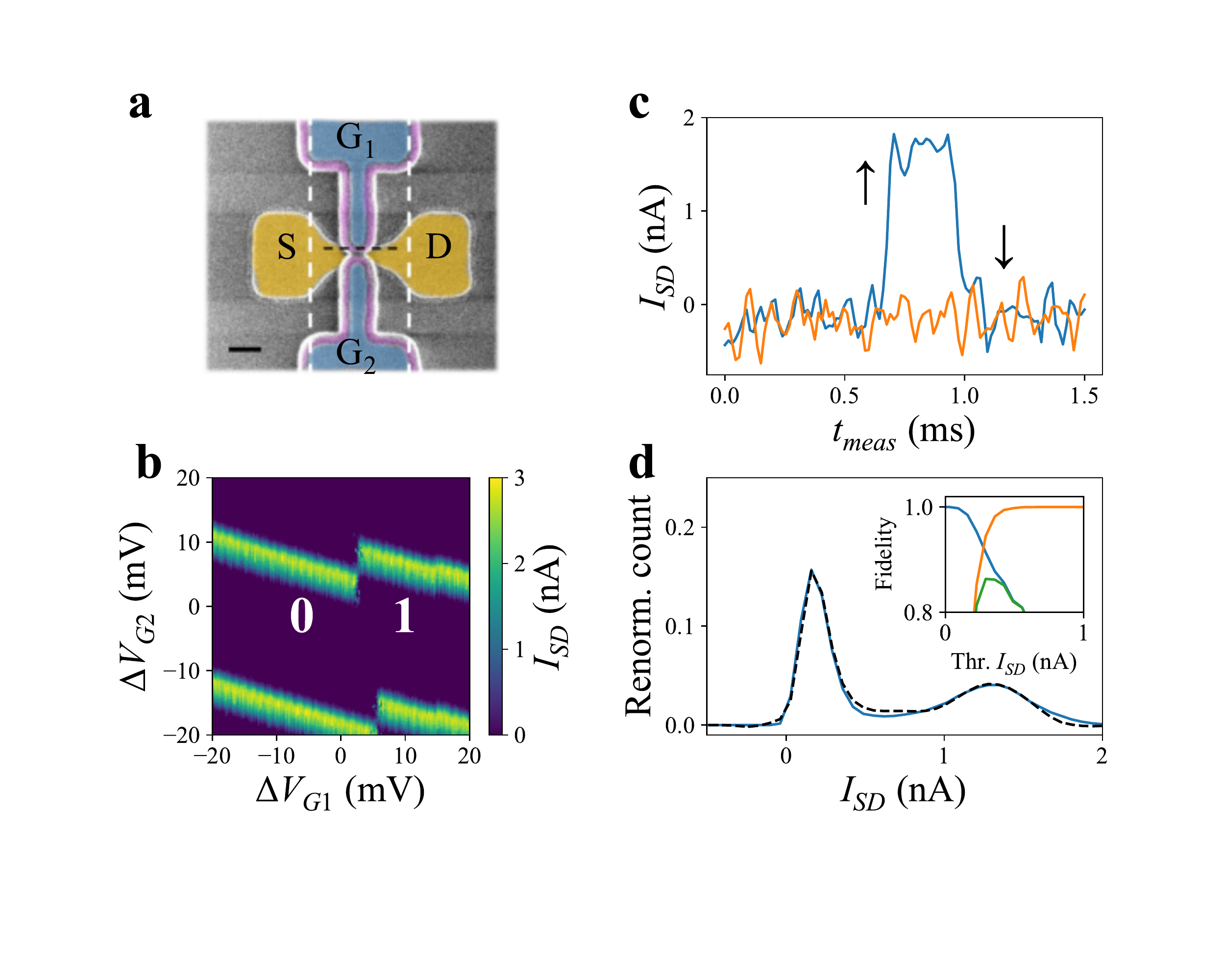}
  \caption{
  (a) Schematic of a CMOS-fabricated nanowire type device identical in structure to the device used here.
The polysilicon channel (yellow) connects two electron reservoirs labelled $S$ and $D$.
The electrostatic environment of the channel is controlled by two front gates, $G_1$ and $G_2$, which are isolated from the channel by $\SI{6}{nm}$ of SiO$_2$ and $\SI{5}{nm}$ of TiN.
A metallic top gate is positioned $\SI{400}{nm}$ above the channel in the region indicated by the white dashed lines and is biased to $+\SI{2}{V}$ to increase the coupling between the dots.
Finally, the silicon bulk below the buried oxide is polarized and used as a back gate(+5V).
(b) Stability diagram of the first electron transition.
The electron occupancy of the qubit dot under $G_1$ is indicated.
The bright regions of current correspond to coulomb peaks where transport is possible through the sensor dot under $G_2$.
The addition of an electron to the qubit dot causes a shift in the potential of the sensor, indicated by the sharp cut. 
(c) Representative time traces of the current through the sensor dot $I_{SD}$ during spin measurement.
If a spin-up electron is loaded, it is able to rapidly tunnel out of the dot, causing a transient shift in the sensor current as the dot is briefly emptied, indicated in blue.
If a spin-down electron is loaded, it remains in the qubit dot and the sensor current does not change, indicated in orange.
A threshold current $I_{thr}$ is defined to distinguish between the two states.
(d) State fidelity analysis at an optimized measurement point.
A histogram of the maximum current of more than $1000$ measurements is binned (blue).
The black dashed line is the total distribution obtained from simulation of more than 10 000 sample traces using experimental parameters.
Inset: The individual state fidelity is calculated for varying threshold current level, with $F(\left | \uparrow \right \rangle)$ in blue and $F(\left | \downarrow \right \rangle)$ in orange.
The product of these, the state visibility $V$, is plotted in green.
}
  \label{fig:device}%	
  \end{figure}

The two devices we present here are depicted in Fig.\ref{fig:device}(a).
They consist of a pair of split front gates, of length $\SI{50}{nm}$ and separated by $\SI{50}{nm}$, which are lying on a silicon nanowire of width $\SI{90}{nm}$ and thickness $\SI{15}{nm}$. 
Two electron reservoirs are formed by ion implantation for the first device and in situ growth of degenerate n-doped silicon for the second one.
They have been measured at low temperature, between $200$ and $350~\si{\milli\kelvin}$.
At this temperature, a QD is formed under each front gate, $G_1$ and $G_2$.
We operate the device as follows: the top quantum dot (QD) is used to trap a single charge and the bottom one as a charge sensor (CS).
Figure \ref{fig:device}(b) shows the stability diagram with respect to $V_{G1}$, $V_{G2}$.
Shifts in the coulomb peaks indicate it is possible to detect the first electron entering the top QD, using the CS.
When the QD is depleted to the few-electron regime, the current passes only through the CS and electrons are loaded onto QD from CS via the inter-dot tunnel barrier.
We operate the device at the transition signalling the addition of the first electron in QD.

%%%%%%%%%%%%%%%%%%%%%%%%%%%%%%%%%%%%%%%%%%%%%%%%%%%%%%%%%%%%%%%%%%%%%%%%
\section{Single shot readout of a single electron spin}

To readout the spin state, we exploit a spin-to-charge conversion based on energy-selective tunneling \cite{morello2010}.
For this purpose, at finite magnetic field, we first load QD with a single electron and then pulse the chemical potential where only the highest spin state can tunnel out of the dot.
Therefore, when the electron is in a up spin state, we observe a signal click, characteristic of the tunneling out of a up spin electron, followed by the tunnelling in of a down spin electron, depicted in Fig. \ref{fig:device}(c) (blue curve). On the opposite, for a spin down, the signal is constant.
It is important to note that similarly to Ref \cite{morello2010}, in both devices, the loading and unloading of QD is achieved through the CS and not through the reservoir.
This single shot detection of the spin state is further analysed by setting the detection threshold at the point of maximum visibility.
To find this point, we build the histogram presented in Fig. \ref{fig:device}(d) where we bin the maximum of each measurement trace which lasts for $1~\si{\milli\second}$ with a sampling rate at $50~\si{\kilo\hertz}$. %\cite{Urdampilleta2019}?
The visibility is obtained by using the same method as in reference \cite{morello2010}, which consists in simulating single shot traces using the extracted tunneling in and out times (respectively $250~\si{\micro\second}$ and $411~\si{\micro\second}$) to describe the current distribution.
We obtain an average readout fidelity above 92\% (and 86\% visibility) limited by the tunneling in and out rates of the single electron \cite{Keith2019}.

\begin{comment}
To further characterize the readout protocol, we investigate the fidelity of initialising the spin state. 
We prepare a state with equal probability in ground and excited states by pulsing the chemical potential below the Fermi energy to load any electron.
We then immediately measure its spin state and we obtain an initial spin-down population of 58\% (instead of 50\%). This discrepancy can be explained by different loading times for the two spin states.
To prepare a ground state we simply load in the position where only the spin down state is below the Fermi sea (similar position to the readout).
In this case, we obtain a spin down population of 92\% (instead of 100\%) which reflects the infidelity of the readout
\end{comment}

\begin{figure}%
\includegraphics[width=\columnwidth]{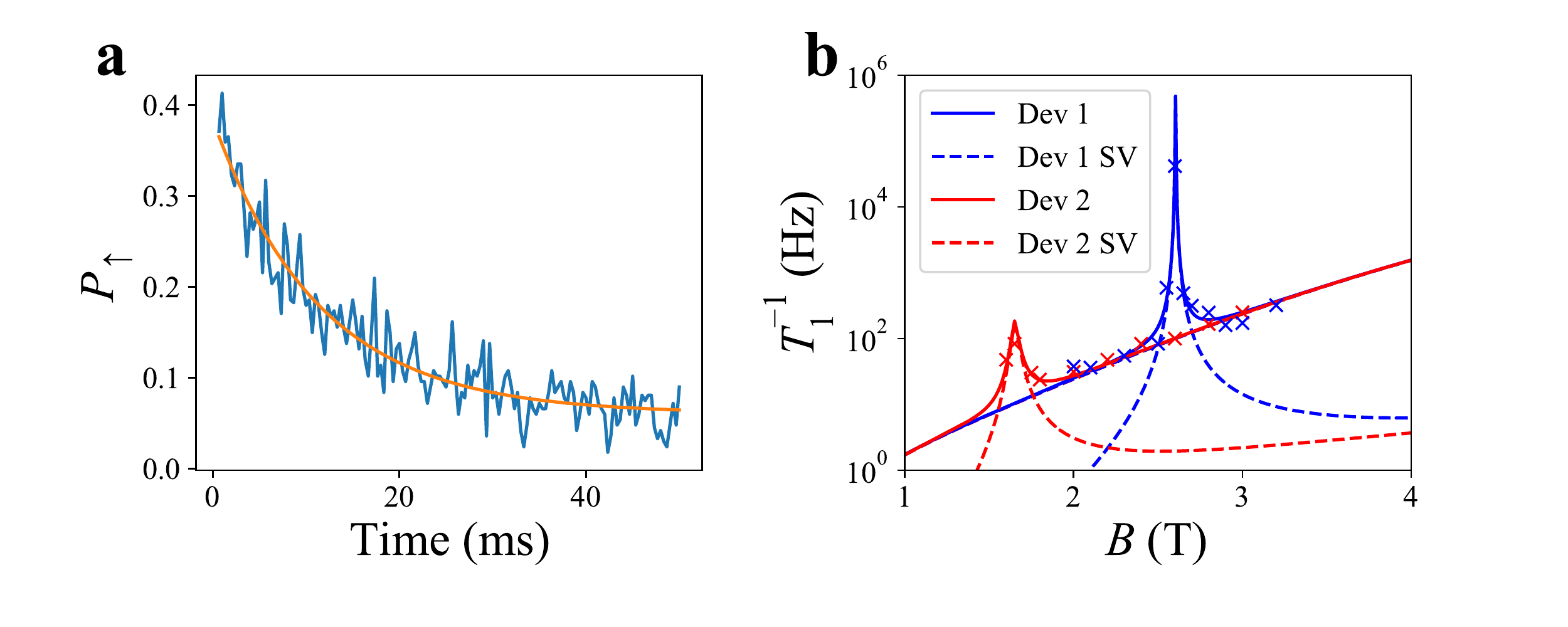}%
\caption{
(a) Spin-up population during measurement as a function of the waiting time after loading.
It is fitted with an exponential decay function to extract $T_1$, the spin relaxation time. 
(b) The average relaxation rate $T_1^{-1}$ for two similar devices is plotted as a function of the magnetic field orthogonal to the nanowire axis.
It is fitted (solid line) with a combination of the spin-valley contribution $\Gamma_{SV}$ (dashed line) and the spin-orbit contributions $\Gamma_{SO}$ to the relaxation rate.
A single prominent increase of relaxation rate, occurring at $\SI{2.6}{T}$ and $\SI{1.7}{T}$ respectively, is induced by spin valley mixing close to $E_{Z} = E_{SV}$ and coupling to the phonon bath.
Outside of this hotspot, the spin relaxation is induced by spin-orbit mixing with higher orbital states.
}
\label{fig:hotspot}%	
\end{figure}

%%%%%%%%%%%%%%%%%%%%%%%%%%%%%%%%%%%%%%%%%%%%%%%%%%%%%%%%%%%%%%%%%%%%%%%%
\section{Spin-valley coupling}
We now move on to the characterization of relaxation time.
Relaxation curves are measured by first loading an electron with random spin orientation, and then probing the spin up population after a given waiting time in the loading region.
We obtain the curve in Fig. \ref{fig:hotspot}(a), which is characteristic of spin relaxation and presents a T$_1$ of $\SI{10}{ms}$.

\subsection{Determination of valley splitting}
To further investigate the presence of excited states, we measure the relaxation rate as a function of the magnetic field on the two devices.
Figure \ref{fig:hotspot}(b) present the results obtained on device 1 and device 2 (blue and red data points respectively).
The two curves feature a hotspot in relaxation at two different magnetic fields ($\SI{2.6}{T}$ and $\SI{1.6}{T}$ respectively).
At these points, the two valley states with opposite spins anticross (the valley splitting $E_{VS}$ is equal to the Zeeman splitting $E_Z$), and give rise to a relaxation channel through spin-valley mixing \cite{huang2014}.

To support this interpretation we fit the data points with a model comprising spin-orbit and spin-valley contributions \cite{huang2014,zhang2020}, which is represented by the blue and red solid lines.
\begin{equation}
  T_1^{-1}=\Gamma_{Ph,SV}+\Gamma_{JN,SO}+\Gamma_{Ph,SO}
\end{equation}
where $\Gamma_{Ph,SV}$ corresponds to the relaxation rate due to spin-valley mixing and coupling to phonons, while $\Gamma_{Ph,SO}$ and $\Gamma_{JN,SO}$ correspond to the relaxation rates due to spin orbit coupling via phonons and Johnson-Nyquist noise respectively.
From these fits we obtain a valley splitting energy, $E_{VS}$, of $\SI{191 \pm 16}{\micro\eV}$ and $\SI{300 \pm13}{\micro\eV}$ respectively and a gap at the spin valley anticrossing of $\SI{4 \pm 0.3}{\micro\eV}$ and $\SI{0.2\pm 0.03}{\micro\eV}$ respectively.
The effect of Johnson-Nyquist noise on the spin-valley contribution has been neglected as at the two hotspots the density of phonon modes is high enough to dominate the relaxation process.
The $\Gamma_{Ph,SV}$ contribution is therefore represented by the dashed blue and red curve which are in good agreement with the experimental data around the hotspot.
The spin orbit contribution (baseline) follows in both cases a similar model which accounts for Johnson-Nyquist noise at low field (where the density of phonons is small) and phonon emission at higher field \cite{huang2014b}.
Interestingly, the two devices present the same relaxation rates outside the hotspots.
This indicates that the overall structure of the quantum dots is similar in both cases.
This is an important result toward the realisation of reliable and low variability quantum dots when operated far from the hotspot.
%%%%%%%%%%%%%%%%%%%%%%%%%%%%%%%%%%%%%%%%%%%%%%%%%%%%%%%%%%%%%%%%%%%%%%%%
\subsection{Anisotropic spin-valley mixing}

\begin{figure}%
  \includegraphics[width=\columnwidth]{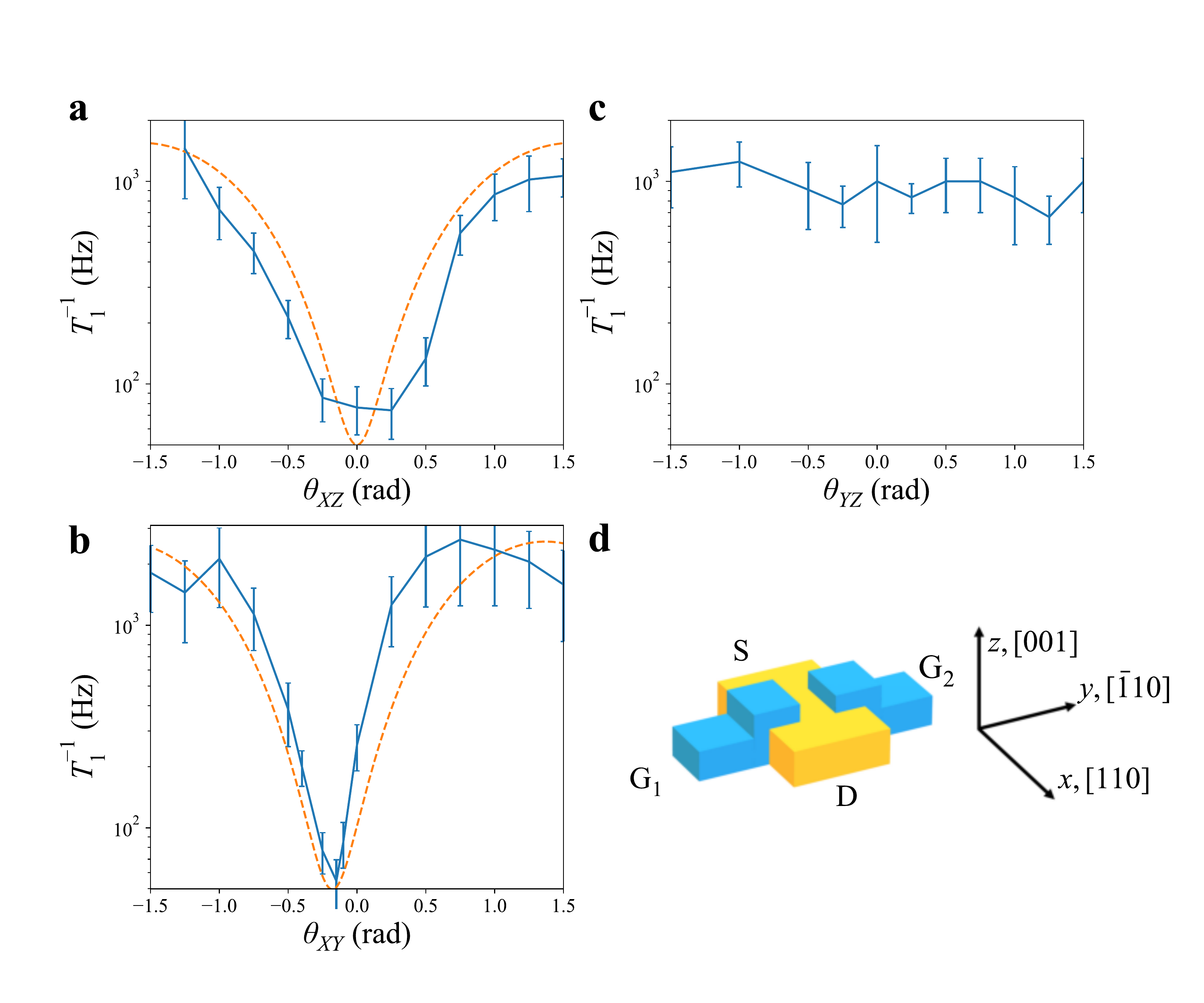}%
  \caption{ Evolution of the relaxation rate with rotation of the magnetic field.
  (a-c) Evolution of the relaxation rate within the $XZ$ ($XY$, $YZ$) plane (under rotation about the $y$ ($z$, $x$) axis, corresponding approximately to the $[\bar{1}10]$ ($[001]$, $[110]$) crystallographic axis).
  Dashed curves in (a) and (b) correspond to sin$^2$ functions fitted to the data.
  (d) Schematic representation of the device and its spatial orientation.
  }
  \label{fig:3D}%	
  \end{figure}

The spin valley mixing is an excellent probe of the local symmetry of the quantum dot \cite{ciriano2021spin, zhang2020, hofmann2017, tanttu2019}.
It is indeed expected that the spin-valley mixing vanishes in the presence of more than one mirror plane in the structure \cite{corna2018}.
The presence of a hotspot is already the sign of a lower symmetry of the system.
The present quantum dots are formed in the corner of the nanowire, therefore, in absence of local disorder, we expect a single mirror plane with its normal vector along $\bold{x}$, the nanowire axis.

To probe the presence of local disorder, we investigate the anisotropic behaviour of the spin-valley mixing.
Such measurements can be used to determine the direction of the spin orbit contribution which can be directly correlated to the local planes of symmetry experienced by the quantum dot.
Figure \ref{fig:3D} (a) (b) and (c) presents the measurement of the relaxation rate as the field direction is rotated in three different orthogonal planes.
To achieve the maximum sensitivity, the magnitude of the magnetic field is set close to the hotspot where the relaxation rate is dominated by the spin-valley mixing ($\SI{2.4}{T}$).
When the magnetic field scans the XZ and YZ plane, we observe a drop in the relaxation rate when it crosses the $x$ axis.
On the opposite the relaxation rate is constant in the YZ plane.

To interpret these data, we need to consider the Hamiltonian $H_{\textrm{SOC}}$ that couples the spin and $v_1$, $v_2$ valley orbitals through the spin-valley mixing matrix element $\bra{v_1\uparrow}H_{\textrm{SOC}}\ket{v_2\downarrow}$, \cite{bourdet2018}.
In the presence of a (yz) mirror symmetry plane $H_{\textrm{SOC}}$ takes the form $H_{\textrm{SOC}}=(\alpha_yp_y + \alpha_zp_z)\sigma_x + (\beta_y\sigma_y + \beta_z\sigma_z)p_x$, with $p_x$, $p_y$ and $p_z$ the electron momentum along \textit{x}, \textit{y} and \textit{z} directions and $\sigma_x$, $\sigma_y$, $\sigma_z$ the Pauli matrices.
The valley orbitals $v_1$ and $v_2$ being invariant by the (yz) mirror plane, $\bra{v_1}p_x\ket{v_2}=\bra{v_1}-p_x\ket{v_2}=0$, see SI of reference \cite{ciriano2021spin}.
Therefore the spin mixing term reads $\bra{v_1\uparrow}(\alpha_yp_y + \alpha_zp_z)\sigma_x \ket{v_2\downarrow}$.
When the magnetic field is aligned with \textit{x}, $\ket{\uparrow}$ and $\ket{\downarrow}$ are eigenstates of $\sigma_x$, which leads to a vanishing spin-valley mixing.
As the magnetic field is tilted away from \textit{x} with an angle $\theta$, the remaining projection leads to $|\bra{v_1\uparrow}H_{\textrm{SOC}}\ket{v_2\downarrow}|^2\propto$ sin$^2\theta$ \cite{corna2018,zhang2020}. 
Following this model, we obtain a quantitative agreement between the experimental data and the sin$^2$ function as shown in Fig. \ref{fig:3D}(a) and (b). 
From these fits, we obtain the minimum relaxation rate along the nanowire axis ([110]), see diagram in Fig. \ref{fig:3D}(d). 
Along this axis, the relaxation is dominated only by the spin-orbit interaction.
It is important to note that there a slight offset in the angle in the XY measurement because the nanowire axis is not perfectly aligned with the coil axes.
Moreover, the small discrepancy between the XY and XZ measurements suggests that Y and Z directions are not equivalent.
This second order anisotropy can be explained by the overlap of the front gate being non equivalent between the side and the top facets of the wire.

%%%%%%%%%%%%%%%%%%%%%%%%%%%%%%%%%%%%%%%%%%%%%%%%%%%%%%%%%%%%%%%%%%%%%%%%
\section{Low frequency charge noise on valley-splitting}
\begin{figure}%
  \includegraphics[width=\columnwidth]{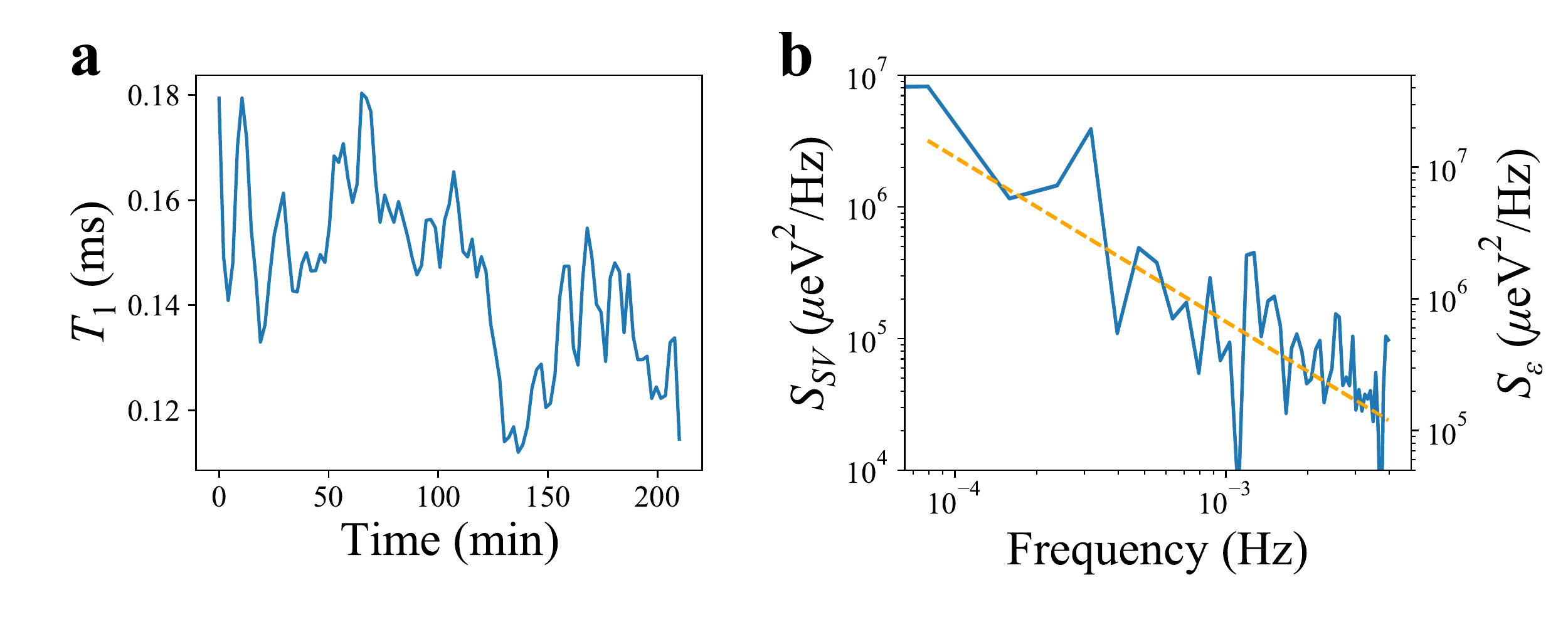}%
  \caption{
(a) Variation in the spin-lattice relaxation rate over time. 
Each measurement point represents a $T_1$ measurement of duration $\SI{4}{min}$. 
(b) Frequency-domain fluctuations of the valley splitting energy $E_{VS}$.
The fit is proportional to $1/f^{1.25}$.
Extrapolation to $\SI{1}{Hz}$ reveals fluctuations in $E_{VS}$ of $\SI{23 \pm 1 }{\micro eV^2 / Hz}$.
%The right axis is scaled to show the power spectral density of the $T_1$ low-frequency noise.
%The PSD is converted to top gate fluctuations using the spin-valley field tuning gradient and back gate lever arm.
%Extrapolation to $\SI{1}{Hz}$ yields a noise level of $\SI{121 \pm 8 }{\micro eV^2 / Hz}$.
}
  \label{fig:noise}%	
\end{figure}

The valley splitting is sensitive to the electric field as it arises from the strong confinement against the top interface \cite{struck2020}.
Here, we use the relaxation at the hotspot to probe the local fluctuations in electric field induced by low frequency charge noise.
For this purpose, we sit at a measurement point next to the hotspot and record the evolution of $T_1$ with time.
To transform this time evolution in a power spectral density for $E_{VS}$, we assume that the dependence of the valley splitting on the electric field is linear for small noise amplitude \cite{ibberson2018}.
From the hotspot fitting we extract the gradient at the measurement point, which combined with a Fourrier transform of the time domain signal, Fig. \ref{fig:noise}(a), yields the power spectral density (PSD) in Fig. \ref{fig:noise}(b).

The PSD follows a $1/f$ dependence which, extrapolated at $\SI{1}{Hz}$, gives $\SI{23}{\micro eV^2/Hz}$.
In term of qubit energy fluctuations, the obtained value is relatively large if we consider the operation of the qubit in a spin-valley mode. 
In this mode, the spin-valley mixing is exploited to drive coherent oscillations through electric dipole spin resonance next to the hotspot \cite{bourdet2018}.
Fluctuations of the valley splitting translate into fluctuations of the Larmor frequency $h\delta f = 1/2\delta E_{VS}$ at the hotspot\cite{corna2018}.
It corresponds to a fluctuation in the spin precession of about $\SI{0.6}{GHz}/\sqrt{Hz}$, which is faster than the decoherence rate due to the hyperfine fluctuations in natural silicon, and is on the order of the decoherence rate of charge and valley qubits \cite{kim2015, Pent2019}. %cite kim2015
This could prove detrimental for EDSR exploiting the spin-valley mode, limiting the coherence time when the spin-valley coupling is turned on \cite{culcer2012, yang2013}.
It may therefore be necessary to operate further away from the hotspot to avoid fast decoherence, at the cost of slower Rabi frequency  or to lower the confinement potential to reduce the valley splitting sensitivity to electric field \cite{bourdet2018}.
\\
\begin{comment}
This value is equivalent to a fluctuation on the quantum dot potential of $\SI{121 \pm 8 }{\micro eV^2/Hz}$, which is extracted based on the valley splitting dependence on the back gate voltage and the respective lever arms.
This is consistent with charge noise amplitude measured via tunnel rate fluctuations in the same device \cite{spence2021}. % cite our charge noise paper?
It is a strong indication that the low-frequency fluctuations which cause the variation in valley splitting energy is charge noise-induced through trapping and de-trapping events at the material interfaces.
\end{comment}

%%%%%%%%%%%%%%%%%%%%%%%%%%%%%%%%%%%%%%%%%%%%%%%%%%%%%%%%%%%%%%%%%%%%%%%%
\section{Conclusion}
In conclusion, we have demonstrated fast and reproducible spin characterization in CMOS nanowire devices.
Similar spin-orbit induced relaxation rate was found across two identically patterned and fabricated devices, which is an important result toward the improvement of the CMOS fabrication process at the quantum dot level.
Moreover, the spin-valley coupling was found to be highly anisotropic, with a strong symmetry plane oriented perpendicular to the channel axis, indicating the absence of strong local disorder at the quantum dot site.
%The sin$^2$ angular dependence is stronger than that obtained in a similar device \cite{ciriano2021spin}.
%It suggests a stronger symmetry plane in the device measured here, indicating lower local disorder near the quantum dot.
These results are of prime importance toward the fabrication of large scale structures containing many quantum dots with low variability.
Finally, low-frequency fluctuations in the valley splitting were measured to cause fluctuations in the Larmor frequency at $\SI{0.6}{GHz}/\sqrt{Hz}$ which would be detrimental to the operation of the qubit as a spin-valley or valley qubit to enhance Rabi frequencies. This motivates further work on dependence and sensitivity of the valley splitting with respect to the electric field and the investigation of potential sweetspot.

\section{Acknowledgments}
We acknowledge fruitful discussions with J. Li and L. Hutin and  support from P. Perrier, H. Rodenas, E. Eyraud, D. Lepoittevin, I. Pheng, T. Crozes, L. Del Rey, D. Dufeu, J. Jarreau, J. Minet and C. Guttin. 
D.J.N. , B.K. and C.S. acknowledge the GreQuE doctoral programs (G.A. No 754303). The device fabrication is funded through the QuCube project (G. A. 810504)
This work is supported by the Agence Nationale de la Recherche through the CRYMCO and MAQSi projects (ANR-21-XXX).

\end{document}